\documentclass[prc,a4paper,onecolumn,amsmath,amssymb,showpacs,showkeys,floatfix]{revtex4}
\usepackage[latin1]{inputenc}
\usepackage[T1]{fontenc}
\usepackage{units}
\usepackage{graphicx}
\usepackage{dcolumn}

\begin{document}

\title{Nuclear matrix elements of $0\nu\beta\beta$ decay with 
improved short-range correlations}
\author{Markus Kortelainen}
\author{Jouni Suhonen}
\affiliation{Department of Physics, University of Jyv\"{a}skyl\"{a}, 
P.O.\ Box 35, FIN-40351 Jyv\"{a}skyl\"{a}, Finland}

\begin{abstract}
Nuclear matrix elements of the neutrinoless double
beta ($0\nu\beta\beta$) decays of $^{96}$Zr, $^{100}$Mo, $^{116}$Cd,
$^{128}$Te, $^{130}$Te and $^{136}$Xe are calculated for the
light-neutrino exchange mechanism by using the proton-neutron quasiparticle
random-phase approximation (pnQRPA) with a realistic nucleon-nucleon force.
The $g_{\rm pp}$ parameter of the pnQRPA is fixed by the data on the
two-neutrino double beta decays and single beta decays.
The finite size of a nucleon, the higher-order terms of nucleonic 
weak currents, and the nucleon-nucleon short-range correlations
(s.r.c) are taken into account. The s.r.c. are computed by the
traditional Jastrow method and by the more advanced unitary
correlation operator method (UCOM). Comparison of the results obtained
by the two methods is carried out. The UCOM computed matrix elements
turn out to be considerably larger than the Jastrow computed
ones. This result is important for the assessment of the neutrino-mass 
sensitivity of the present and future double beta experiments.
\end{abstract}


\maketitle

\section{Introduction}

The neutrinoless double beta ($0\nu\beta\beta$) decay plays a key role
in the search for massive Majorana neutrinos and their mass
scale. The experimental search for $0\nu\beta\beta$ decay has become 
front-line physics due to the verification of the existence of
neutrino mass by the oscillation experiments \cite{Oscillations} and 
the claimed discovery of the $0\nu\beta\beta$ decay \cite{KLA01,KLA04}.
At present there are two important experiments taking data, namely the NEMO 3 
\cite{ARN05a} and CUORICINO \cite{ARN05b} experiments. A host of
important future experiments are under R\&D planning and 
construction \cite{BAR04}. For all these expensive experiments the
computed values of the involved nuclear matrix elements have become 
an important issue \cite{ELL02,ELL04,BIL04,GEH07}. They are
essential when one starts to extract quantitative neutrino properties
from the measured data.

Many nuclear models of different types have been deviced to compute
the nuclear matrix elements of the $0\nu\beta\beta$ decay
\cite{SUH98}. There are essentially two complementary families of nuclear
models on the market, the nuclear shell model \cite{HAX84,CAU99,CAU05} and
the proton-neutron quasiparticle random-phase approximation
(pnQRPA) \cite{VOG86,CIV87,SUH98}. The pnQRPA is constructed to describe the 
energy levels of odd-odd nuclei and their beta decays to the
neighboring even-even nuclei \cite{SUH07}. Also its derivative,
renormalized pnQRPA \cite{TOI95}, has been used \cite{TOI97,ROD06} to
compute double-beta matrix elements. The pnQRPA (and the renormalized pnQRPA) 
calculations can be fine tuned by the so-called 
particle-particle strength parameter, $g_{\rm pp}$, that
controls the magnitude of the proton-neutron two-body interaction 
for the $1^+$ intermediate states in double beta decay \cite{VOG86,CIV87}.
There are two ways to fix the value of this parameter, either by using
the data on two-neutrino double beta ($2\nu\beta\beta$) decays
\cite{ROD06} or the data on single beta decays \cite{SUH05,CIV05}.
In this work we use the $2\nu\beta\beta$ data to fix the possible
values of $g_{\rm pp}$ and cross check it by single beta decays
wherever possible.

In the mass mode of the $0\nu\beta\beta$ decay
a light virtual Majorana neutrino is exchanged by the two decaying
neutrons of the initial nucleus. The average exchanged momentum is
large so that the two neutrons tend to overlap. To prevent this
a Jastrow type of correlation function was introduced in
\cite{HAX84,ENG88} following the parametrization of Miller and
Spencer \cite{MIL76}. This method, although microscopically inspired,
is just a phenomenological way to introduce \emph{short-range correlations}
into the two-nucleon relative wave function. The Jastrow function
simply cuts off the wave function of the two nucleons at short
relative distances $r$ leading to a violation of the norm of the wave function.

In the present calculations we improve on the Jastrow method by engaging
the more sophisticated microscopic approach of unitary correlation operator 
method (UCOM) \cite{FEL98}. In the UCOM one obtains the correlated 
many-particle state from the uncorrelated one by a unitary
transformation and thus the norm of the correlated state is 
conserved and no amplitude is lost in the relative wave function. 
In the $0\nu\beta\beta$ calculations this leads to a more complete
description of the relative wave function for small
$r$, as was demonstrated in \cite{KOR07} for the decays of 
$^{48}$Ca and $^{76}$Ge. In this work and in \cite{KOR07b}, where the
decays of $^{76}$Ge and $^{82}$Se were analyzed,
it is demonstrated that the Jastrow procedure leads to the 
excessive reduction of 25\% -- 40\% in the magnitudes of the 
$0\nu\beta\beta$ nuclear matrix elements. At the same time the UCOM 
reduces the magnitudes of the matrix elements only by 4\% -- 16\%. 
The magnitude of the short-range corrections affects the magnitudes of
the nuclear matrix elements which, in turn, dictate the neutrino-mass
sensitivity of the potentially succesful future double beta experiments.
The notable differences between the Jastrow and UCOM corrections
influence the cost estimates of large-scale experiments if a
given neutrino-mass sensitivity is wanted.

In the article we continue the work of \cite{KOR07b}, where the
$0\nu\beta\beta$ nuclear matrix elements of $^{76}$Ge 
and $^{82}$Se were derived. We apply the UCOM and Jastrow
short-range correlations on matrix elements derived by the pnQRPA
method and corrected for the 
\emph{higher-order terms of nucleonic weak current} and the
\emph{nucleon's finite size} using the recipes of \cite{ROD06,SIM99}.
We analyze the $0\nu\beta\beta$-decay matrix elements of 
$^{96}$Zr, $^{100}$Mo, $^{116}$Cd, $^{128}$Te, $^{130}$Te and
$^{136}$Xe for all the mentioned corrections. The necessary
theoretical background is briefly described
in Sec. II and the numerical application is reviewed in Sec. III. The
results are discussed in Sec. IV, and the summary and conclusions are
presented in Sec. V.

\section{Theoretical background}

We start this short review of the theory by presenting the expression
for the half-life of the $2\nu\beta\beta$ decay:
\begin{equation}
\left\lbrack t_{1/2}^{(2 \nu)}(0_{\rm i}^+ \rightarrow 0_{\rm f}^+)
\right\rbrack^{-1}=
G^{(2\nu )}\left\vert M_{\rm DGT}^{(2\nu )}\right\vert^2 \ .
\label{eq:2vbb}
\end{equation}
The transition proceeds from the initial ground state, $0^+_{\rm i}$, to the 
final ground state, $0^+_{\rm f}$. Here $ G^{(2 \nu)}$  is an integral
over the phase space of the leptonic variables \cite{SUH98,DOI85}.
The involved double Gamow--Teller matrix element, $M_{\rm {DGT}}^{(2 \nu)}$, 
can be written as
\begin{eqnarray} 
M_{ \rm {DGT}}^{(2 \nu)} & = & \sum_{n} \frac{
(0_{\rm f}^{+} \mid \mid \sum_j \mbox{\boldmath{$\sigma$}}(j) t^{-}_j
\mid \mid 1_n^{+})}{( {{1} \over {2}} Q_{\beta \beta}+
E_n -M_{\rm i})/ m_{\rm e} +1} \nonumber \\
 & & \times (1_n^{+} \mid \mid \sum_j \mbox{\boldmath{$\sigma$}}(j) t^{-}_j
\mid \mid 0_{\rm i}^+)  \; ,
\label{eq:mdgt}
\end{eqnarray}
where the transition operators are the usual Gamow-Teller operators for $\beta^-$
transitions, $Q_{\beta \beta}$ is the $2\nu\beta\beta$ $Q$ value, $E_n$ is the
energy of the $n$th intermediate state, $M_{\rm i}$ is the mass energy
of the initial nucleus, and $m_{\rm e}$ is the rest-mass of the
electron. It has to be noted here that the expression (\ref{eq:mdgt})
is scaled by the electron rest mass to yield a dimensionless matrix
element. This definition deviates from that of some other authors,
like \cite{ROD06}, where the scaling is not done.

The $0\nu\beta\beta$ decay can proceed via the exchange of a
light virtual Majorana neutrino. Assuming this neutrino-mass 
mechanism to be the dominant one, 
the inverse of the $0\nu\beta\beta$ half-life can be written as
\begin{equation} \label{eq:0nbbhl}
\left[ t_{1/2}^{(0\nu)}(0_{\rm i}^+ \rightarrow 0_{\rm f}^+)
\right]^{-1} = G_{1}^{(0\nu)}
\left( \frac{\langle m_{\nu}\rangle }{m_{\rm e}} \right)^{2}
\left( M^{(0\nu)}\right)^{2}\ ,\quad M^{(0\nu)} = M_{\rm GT}^{(0\nu)} - 
\left( \frac{g_{\rm V}}{g_{\rm A}}
\right)^{2} M_{\rm F}^{(0\nu)} + M_{\rm T}^{(0\nu)} .
\end{equation}
In the above expression $M^{(0\nu)}$ is the total nuclear matrix
element consisting of the Fermi, Gamow--Teller and tensor
contributions. The effective mass of the neutrino is given by
\begin{equation}
\langle m_{\nu}\rangle = \sum_{j} \lambda^{\rm CP}_{j} m_{j}
|U_{{\rm e}j}|^{2} \ ,
\end{equation}
where $\lambda^{\rm CP}_{j}$
is the CP phase and $U_{{\rm e}j}$ is a component of the
neutrino mixing matrix. The definition of the leptonic 
phase-space factor $G_{1}^{(0\nu)}$ can be found in \cite{SUH98}.

The nuclear matrix elements involved in the mass mode of the
$0\nu\beta\beta$ decay are defined as
\begin{eqnarray} \label{eq:bbfermi}
M_{\rm F}^{(0\nu)} & = & \sum_{a} (0^{+}_{\rm f} || h_{\rm F}
(r_{mn},E_{a}) || 0^{+}_{\rm i}) \, , \\
\label{eq:bbgamowt}
M_{\rm GT}^{(0\nu)} & = & \sum_{a} (0^{+}_{\rm f} || h_{\rm GT}
(r_{mn},E_{a})  \mbox{\boldmath{$\sigma$}}_{m}\cdot 
\mbox{\boldmath{$\sigma$}}_{n}  || 0^{+}_{\rm i}) \, ,
\end{eqnarray}
where the summation 
runs over all the intermediate states and the integration is taken
over the relative coordinate $r_{mn}=\vert\mathbf{r}_m -
\mathbf{r}_n\vert$ between the nucleons $m$ and $n$. The neutrino potential
$h_{K}(r_{mn},E_{a})$, $K={\rm F,GT}$, is defined as
\begin{equation} \label{eq:nupot}
h_{K}(r_{mn},E_{a}) = \frac{2}{\pi}R_{\rm A} \int dq
\frac{qh_{K}(q^{2})}{q+E_{a}-(E_{\rm i}+E_{\rm f})/2} j_{0}(qr_{mn}) \, ,
\end{equation}
where $R_{A} = 1.2A^{1/3}\,{\rm fm}$ is the nuclear radius and 
$j_{0}$ is the spherical Bessel function. The term 
$h_{K}(q^{2})$ in (\ref{eq:nupot}) includes the contributions arising
from the induced currents and the finite nucleon size \cite{ROD06,SIM99}.

Next we write the nuclear matrix elements explicitly in the pnQRPA 
framework. They are given by
\begin{eqnarray}
M_{K}^{(0\nu)} & = & \sum_{J^{\pi},k_{1},k_{2},J'} \sum_{pp'nn'}
(-1)^{j_{n}+j_{p'}+J+J'} \sqrt{2J'+1} \nonumber \\
& & \times \left\{ \begin{array}{ccc} j_{p} & j_{n} & J \\
j_{n'} & j_{p'} & J'\end{array} \right\} 
( pp':J' \vert\vert  {\mathcal O}_K \vert\vert nn':J' )
\\ & & \times
( 0^{+}_{\rm f} \vert \vert  \left[ c^{\dag}_{p'}
\tilde{c}_{n'}\right]_J \vert \vert J^{\pi}_{k_{1}} )
\langle J^{\pi}_{k_{1}} \vert J^{\pi}_{k_{2}} \rangle
( J^{\pi}_{k_{2}} \vert \vert  \left[ c^{\dag}_{p}
\tilde{c}_{n}\right]_J \vert \vert 0^{+}_{\rm i}) \, , \nonumber
\end{eqnarray}
where $k_1$ and $k_2$ label the different pnQRPA solutions for a given
multipole $J^{\pi}$. The operators ${\mathcal O}_K$ inside the
two-particle matrix element derive from (\ref{eq:bbfermi}) and
(\ref{eq:bbgamowt}) and they can be written as
\begin{equation}
{\mathcal O}_{\rm F} = h_{\rm F}(r,E_k)\ ,\quad 
{\mathcal O}_{\rm GT} = h_{\rm GT}(r,E_k)\mbox{\boldmath{$\sigma$}}_{1}\cdot 
\mbox{\boldmath{$\sigma$}}_{2} \ ,\quad r=\vert\mathbf{r}_1 -
\mathbf{r}_2\vert \ .
\end{equation}
The expression for the pnQRPA transition densities 
$( 0^{+}_{\rm f} \vert \vert  [ c^{\dag}_{p'}
\tilde{c}_{n'}]_J \vert \vert J^{\pi}_{k_{1}} )$ and
$( J^{\pi}_{k_{2}} \vert \vert  [ c^{\dag}_{p}
\tilde{c}_{n}]_J \vert \vert 0^{+}_{\rm i})$, and
for the overlap factor 
$\langle J^{\pi}_{k_{1}} \vert J^{\pi}_{k_{2}} \rangle$
can be found e.g. in \cite{SUH98,SUH07}.

The traditional way to include the short-range correlations in the
$0\nu\beta\beta$ decay calculations is by introducing the Jastrow
correlation function $f_{\rm J}(r)$, which depends on the relative
distance $r=|{\bf r}_{1}-{\bf r}_{2}|$ of two nucleons. In the Jastrow
scheme the uncorrelated operator ${\mathcal O}$ is replaced by the 
correlated operator ${\mathcal O}_{\rm J}$ by a procedure
\begin{equation} \label{eq:Osrc}
(0^{+}_{f} || {\mathcal O} || 0^{+}_{i}) \to
(0^{+}_{f} || {\mathcal O}_{\rm J} || 0^{+}_{i}) =
(0^{+}_{f} || f_{\rm J} {\mathcal O} f_{\rm J} || 0^{+}_{i}) \, .
\end{equation}
A typical choice for the function $f_{\rm J}$ in $0\nu\beta\beta$
calculations is \cite{TOM91}
\begin{equation} \label{eq:jastrow}
f_{\rm J}(r) = 1 - e^{-ar}\left( 1- br^{2} \right) \, ,
\end{equation}
with $a=1.1 \, {\rm fm}^{2}$ and $b=0.68 \, {\rm fm}^{2}$.
Such application of the Jastrow correlation function is very
rudimentary and, consequently, the Jastrow correlation cuts out the
$r\leq 1\,\textrm{fm}$ part from the relative two-particle wave 
function. In Ref. \cite{KOR07} it was demonstrated that this leads 
to overestimation of the effects of short-range correlations on
the many-body wave function. 

To circumvent the difficulties associated to the Jastrow correlations 
the more refined unitary correlation operator method (UCOM) 
\cite{FEL98} was used in the $0\nu\beta\beta$ decay calculations 
of Ref. \cite{KOR07}. In UCOM one obtains the correlated many-body 
state $\vert \tilde{\Psi} \rangle$ from the uncorrelated one as
\begin{equation}
\vert \tilde{\Psi} \rangle = C \vert\Psi\rangle \, ,
\end{equation}
where $C$ is the unitary correlation operator. The operator $C$
is a product of two unitary operators: $C=C_{\Omega}C_{r}$, where
$C_{\Omega}$ describes short-range tensor correlations and
$C_{r}$ central correlations. Due to the unitary of the operator
$C$ the norm of the correlated state is conserved. Moreover, since 
the correlated matrix element of the operator ${\mathcal O}$ can be 
written as
\begin{equation}
\langle \tilde{\Psi} \vert {\mathcal O} \vert \tilde{\Psi}' \rangle
= \langle \Psi \vert C^{\dag}{\mathcal O} C \vert \Psi' \rangle
= \langle \Psi \vert \tilde{{\mathcal O}}  \vert \Psi' \rangle ,
\end{equation}
it is therefore equivalent to use either correlated states or
correlated operators. For the Fermi and Gamow-Teller
$0\nu\beta\beta$ nuclear matrix elements the effect of the
tensor correlation operator $C_{\Omega}$ vanishes and one is thus
left only with the central correlation operator. The UCOM 
parameters used in our $0\nu\beta\beta$ calculations are 
the Bonn-A parameters taken from \cite{NEF03}.

\section{Numerical application}

In \cite{KOR07b} we applied the pnQRPA to compute the $0\nu\beta\beta$ 
nuclear matrix elements of $^{76}$Ge and $^{82}$Se in the model 
space containing the 1p-0f-2s-1d-0g-0h$_{11/2}$ single-particle orbitals, both for
protons and neutrons. Here we add to this model space the spin-orbit
partner 0h$_{9/2}$ of the 0h$_{11/2}$ orbital, both for protons and
neutrons, to describe the decays of $^{96}$Zr and $^{100}$Mo. For the 
rest of the decays, namely for the $^{116}$Cd, $^{128}$Te, $^{130}$Te
and $^{136}$Xe decays we have extended the proton and neutron
model spaces to include the 1p-0f-2s-1d-0g-2p-1f-0h single-particle
orbitals. The single-particle energies were obtained from
a spherical Coulomb-corrected Woods--Saxon potential with a standard
parametrization, optimized for nuclei near the line of beta
stability. Slight adjustments were done for some of the
energies at the vicinity of the proton and neutron Fermi
surfaces to reproduce better the low-energy spectra of the neighboring
odd-$A$ nuclei and those of the intermediate nuclei.

We have used the Bonn-A G-matrix as the two-body interaction and we have 
renormalized it in the standard way \cite{SUH88,SUH93} by fitting the
pairing parameters of the BCS by comparing with
the phenomenological pairing gaps, extracted
from the atomic mass tables. The particle-hole parameter $g_{\rm ph}$
of the pnQRPA affects the position of the giant Gamow-Teller resonance
and its value was fixed by the available data on the location of the 
giant state. Due to this phenomenological renormalization of the
two-body interaction we did not perform an additional UCOM 
renormalization \cite{ROT05}.

After fixing all the Hamiltonian parameters the
only free parameter left is the proton-neutron particle-particle
parameter $g_{\rm pp}$ of the pnQRPA.
We obtained the physical values of $g_{\rm pp}$ by using the
method of \cite{ROD06,KOR07b}. Consequently, we used the extracted
experimental matrix elements of \cite{ROD06} that include the 
experimental error limits and the
uncertainty in the value of the axial-vector coupling constant 
$1.0\le g_{\rm A}\le 1.25$. The resulting intervals of the
experimental $2\nu\beta\beta$ matrix elements are shown in the second
column of Table~\ref{tab:intervals}, and they are scaled by the 
electron rest mass according to (\ref{eq:mdgt}). 
By performing the pnQRPA calculations of the $2\nu\beta\beta$ matrix
elements the ranges of experimental
matrix elements were subsequently converted to the
intervals of $g_{\rm pp}$ values shown in column three of
Table~\ref{tab:intervals}. For some cases there exists $\log ft$ data on
$\beta^-$ decay from the first $1^+$ state of the intermediate nucleus
to the ground state of the double beta daughter nucleus.
Applying the above procedure to this data leads to the
experimental $\beta^-$ matrix elements and the corresponding ranges of
$g_{\rm pp}$, listed in the last column of Table~\ref{tab:intervals}.

\begin{table}
\caption{\label{tab:intervals} Values of the $g_{\rm pp}$ parameter
  extracted from the data. First column shows the decay and the
  second column the matrix element values extracted from the 
  $2\nu\beta\beta$-decay data by \cite{ROD06}. Third column gives 
  the range of $g_{\rm pp}$ corresponding to the matrix elements of the second
  column. The $g_{\rm pp}$ ranges of the last column were extracted
  from the available $\beta^-$ decay data.}
\begin{ruledtabular}
	\begin{tabular}{cccc}
Decay	& m.e.($2\nu\beta\beta$) & $g_{\rm pp}$($2\nu\beta\beta$) & 
$g_{\rm pp}$($\beta^-$) \\
	\hline
$^{96}\textrm{Zr}\to\,^{96}\textrm{Mo}$ & $0.026-0.112$ & $1.06-1.11$
& - \\
$^{100}\textrm{Mo}\to\,^{100}\textrm{Ru}$ & $0.107-0.181$ & $1.07-1.09$
& $1.07-1.08$ \\
$^{116}\textrm{Cd}\to\,^{116}\textrm{Sn}$ & $0.058-0.102$ & $0.97-1.01$
& $0.82-0.84$ \\
$^{128}\textrm{Te}\to\,^{128}\textrm{Xe}$ & $0.011-0.037$ & $0.89-0.92$
& $0.86-0.88$ \\
$^{130}\textrm{Te}\to\,^{130}\textrm{Xe}$ & $0.014-0.054$ & $0.84-0.90$
& - \\
$^{136}\textrm{Xe}\to\,^{136}\textrm{Ba}$ & $\le 0.023$ & $\ge 0.74$
& - \\
	\end{tabular}
\end{ruledtabular}
\end{table}

As can be seen from Table~\ref{tab:intervals}, the values of 
$g_{\rm pp}$ extracted from the $\beta^-$ and $2\nu\beta\beta$ data
are (roughly) compatible for the decays of $^{100}$Mo and $^{128}$Te,
whereas for the $^{116}$Cd decay this is not the case. 
This discrepancy was already pointed out in \cite{SUH05}. 
The reason for this discrepancy is not clear but an interesting 
observation is that the 
$^{116}$Cd decay obeys very closely the single-state dominance
hypothesis whereas the $^{128}$Te decay obeys it slightly less and
the $^{100}$Mo decay the least, as clearly shown in Table~1 of
\cite{CIV99}.

\section{Discussion of the results}

In Table \ref{t:mcomparison} we show the evolution of the values of the
$0\nu\beta\beta$ nuclear matrix elements as we add more corrections
to the bare matrix element. In the first column we list the used
$g_{\rm pp}$ value which was taken to be in the middle of the
$g_{\rm pp}$ interval of Table~\ref{tab:intervals}. The
following columns list the bare matrix element (b.m.e.), the
matrix element including the higher-order terms of the nucleonic weak
current (b.m.e.+A), and the matrix element with finite nucleon size effects
added (b.m.e.+A+B). In the last two columns we have added
either the Jastrow (C) or UCOM (D) short range correlations.
The value $g_{\rm A}=1.25$ was used in these calculations.

\begin{table}
\caption{Total matrix element $M^{(0\nu)}$ of (\ref{eq:0nbbhl}) 
computed by correcting the bare matrix element (b.m.e) for the 
higher-order terms of the nucleonic weak current (A), for the
finite nucleon size (B), and for either the Jastrow (C) or UCOM (D)
correlations. The used value of $g_{\rm pp}$ is indicated
in the first column.}
\label{t:mcomparison}
\begin{ruledtabular}
\begin{tabular}{lrrrrrr}
Nucleus & $g_{\rm pp}$ &  b.m.e. & $+$A & $+$A$+$B & $+$A$+$B$+$C &
$+$A$+$B$+$D \\
\hline
$^{96}$Zr  & 1.085 & $-5.308$ & $-4.814$ & $-3.736$ & $-2.454$ & $-3.521$ \\
$^{100}$Mo & 1.08  & $-6.126$ & $-5.571$ & $-4.358$ & $-2.914$ & $-4.113$ \\
$^{116}$Cd & 0.99  & $-5.726$ & $-5.172$ & $-4.263$ & $-3.169$ & $-4.076$ \\
$^{128}$Te & 0.905 & $-7.349$ & $-6.673$ & $-5.260$ & $-3.563$ & $-4.979$ \\
$^{130}$Te & 0.87  & $-6.626$ & $-6.021$ & $-4.777$ & $-3.285$ & $-4.530$ \\
$^{136}$Xe & 0.74  & $-4.715$ & $-4.269$ & $-3.478$ & $-2.537$ & $-3.317$ \\
\end{tabular}
\end{ruledtabular}
\end{table}

Table \ref{t:mcomparison} shows that the Jastrow method
produces a much larger reduction in the magnitude of 
$M^{(0\nu)}$ than the UCOM. The UCOM (D) with higher-order term (A) and
finite nucleon size (B) corrections included seems to reduce the
magnitude of $M^{(0\nu)}$ by a rough factor of $2/3$ from its bare value.
A similar scaling factor was also present for the $0\nu\beta\beta$ results
of $^{76}$Ge and $^{82}$Se in \cite{KOR07b}. At the same time 
one obtains a reduction factor of $1/2$ for the corresponding
Jastrow (C) results.

\begin{figure} \begin{center}
\includegraphics[width=10cm]{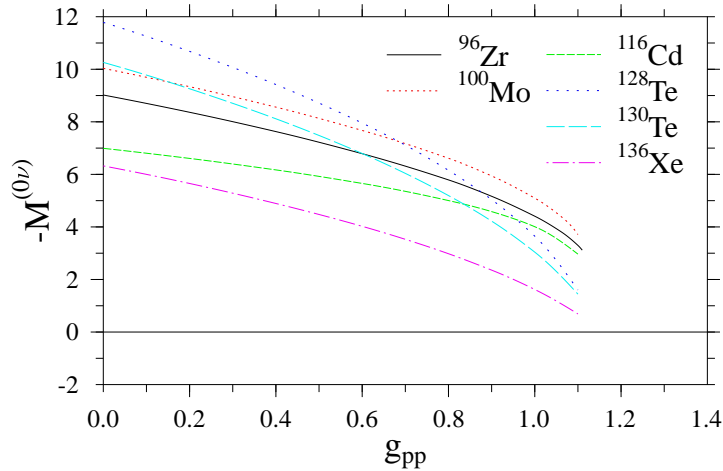} \end{center}
\caption{(Color online) Calculated values of $M^{(0\nu)}$ for the indicated
  $0\nu\beta\beta$ decays as functions of $g_{\rm pp}$. The UCOM (D) was
  used with $g_{\rm A}=1.25$ including all the other (+A+B) corrections.}
\label{f:mgpp}
\end{figure}

We visualize the $g_{\rm pp}$ dependence of $M^{(0\nu)}$ in
Fig.~\ref{f:mgpp}.  The calculations included the higher-order
term (A), finite-size (B) and UCOM (D) corrections. The calculations
are shown for the interval $g_{\rm pp}\le 1.1$, the upper limit lying 
near the breaking point of the pnQRPA for all shown nuclear systems. 

\begin{figure} \begin{center}
\includegraphics[width=15cm]{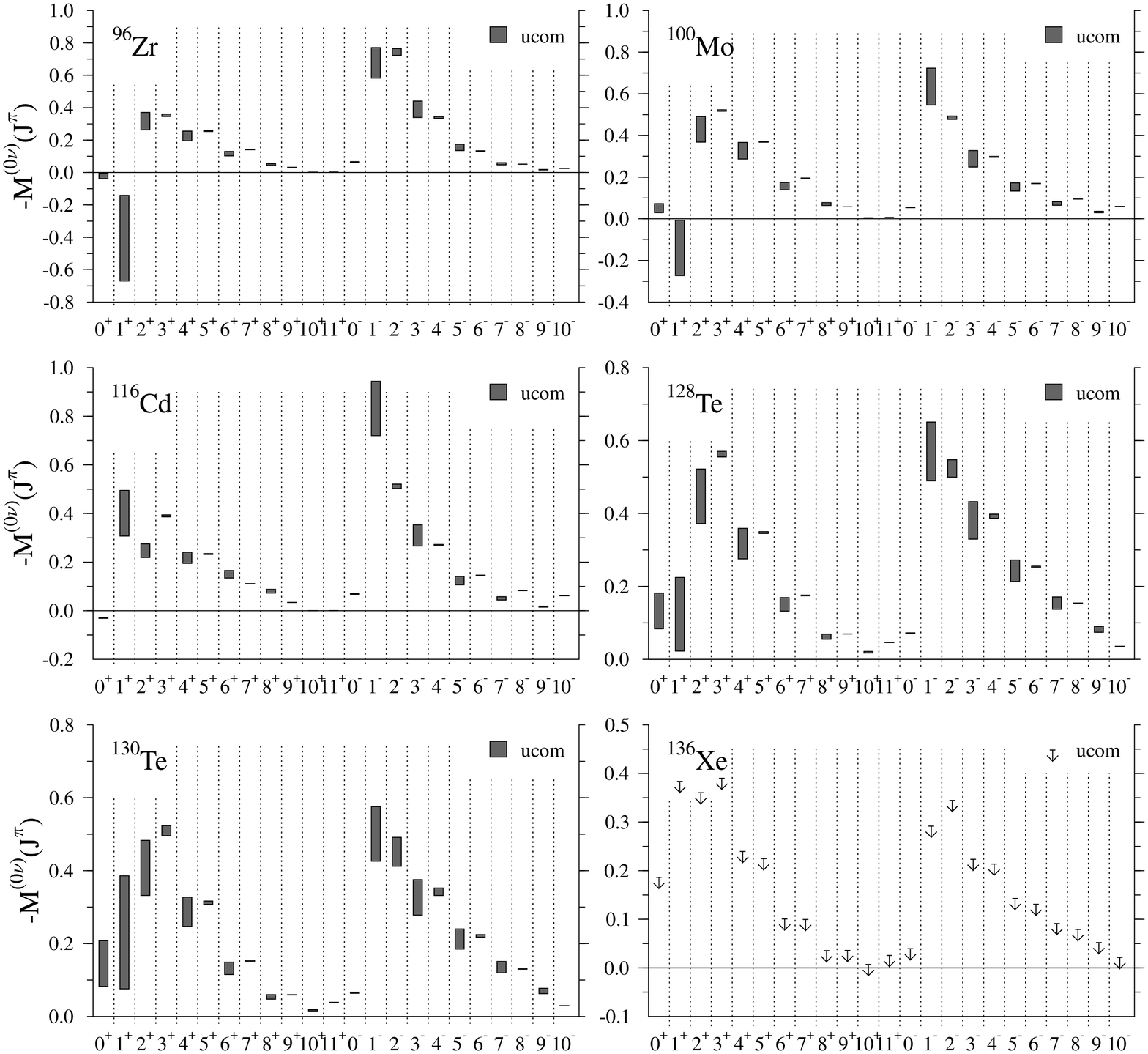} \end{center}
\caption{Multipole decomposition of $M^{(0\nu)}$ for the calculated
  $0\nu\beta\beta$ decays.}
\label{f:multipoles}
\end{figure}

In Fig. \ref{f:multipoles} we have plotted the multipole decomposition
of the total $0\nu\beta\beta$ matrix element $M^{(0\nu)}$ for all 
calculated decays. The upper (lower) end of each bar corresponds to
the lower (upper) end of the $g_{\rm pp}$ interval of
Table~\ref{tab:intervals}. For the $^{136}$Xe decay we can only give 
upper limits since only the lower limit of $g_{\rm pp}$ is known as
indicated in the last line of Table~\ref{tab:intervals}. 
From the Fig. \ref{f:multipoles} one can see that the 
widest spread appears in the bar corresponding to the $1^{+}$
contribution. Furthermore, the $g_{\rm pp}$ interval extracted from
the $2\nu\beta\beta$ data confines the $1^{+}$ contribution in a
striking way: for $^{96}$Zr and $^{100}$Mo the $1^{+}$ contribution is
of opposite sign to the other contributions. This interference with
the rest of the contributions reduces the magnitude of $M^{(0\nu)}$
for these two decays. Another notable feature is that the $1^-$
contribution is always the leading one, the $2^-$ contribution being
usually of comparable size. This pattern is different from the one of
$^{76}$Ge and $^{82}$Se decays where the $2^-$ contribution was the
dominant one \cite{KOR07}. All these observations are in
qualitative agreement with the results of \cite{ROD06}.

We have collected the obtained $g_{\rm pp}$ limits of
Table~\ref{tab:intervals} and their 
corresponding $0\nu\beta\beta$ matrix elements of (\ref{eq:0nbbhl}) in
Table~\ref{t:mvali}. It is worth pointing out that the magnitudes of the
tensor matrix element $M_{\rm T}^{(0\nu)}$ are
quite small and are not indicated in the table.
As discussed earlier, the $g_{\rm pp}$ limits
arise from both the $2\nu\beta\beta$ and beta decay data.
For $^{136}$Xe only experimental lower limit of the $2\nu\beta\beta$ 
half-life is available, yielding only a lower limit for $g_{\rm pp}$. For
$^{128}$Te the available beta decay data give $g_{\rm pp}=0.86$ 
and for $^{116}$Cd they give $g_{\rm pp}=0.82$. In the last column of
Table~\ref{t:mvali} we also tabulate our predicted half-life limits in 
units of ${\rm yr}/(\langle m_{\nu} \rangle [{\rm eV}])^{2}$. The
beta-decay data yields a different $g_{\rm pp}$ interval than the data
on $2\nu\beta\beta$ decays for $^{116}$Cd and $^{128}$Te. This
discrepancy is especially striking for $^{116}$Cd. Implications of
this discrepancy and its cure are still open questions \cite{SUH05}.

\begin{table}
\caption{Calculated $0\nu\beta\beta$ nuclear matrix elements, the 
used $g_{\rm pp}$ and $g_{\rm A}$ values and the resulting half-lives. 
The UCOM and other corrections are included. The half-lives 
$t_{1/2}^{(0\nu)}$ are expressed in units of 
${\rm yr}/(\langle m_{\nu} \rangle [{\rm eV}])^{2}$.}
\label{t:mvali}
\begin{ruledtabular}
\begin{tabular}{ccccccc}
Nucleus & $g_{\rm pp}$ & $g_{\rm A}$ & $M_{\rm F}^{(0\nu)}$ & 
$M_{\rm GT}^{(0\nu)}$ & $M^{(0\nu)}$ & $t^{(0\nu)}_{1/2}$ \\
\hline
$^{96}$Zr  & 1.06 & 1.00 & 1.350 & $-$2.969 & $-$4.319 & $6.1\times 10^{23}$ \\
           & 1.11 & 1.25 & 1.261 & $-$2.315 & $-$3.117 & $4.7\times 10^{23}$ \\
\hline
$^{100}$Mo & 1.07 & 1.00 & 1.583 & $-$3.266 & $-$4.849 & $6.2\times 10^{23}$\\
           & 1.09 & 1.25 & 1.543 & $-$2.950 & $-$3.931 & $3.8\times 10^{23}$\\
\hline
$^{116}$Cd & 0.82 ($\beta^-$ decay) & 1.25 & 1.427 & $-$4.021 &
$-$4.928 & $2.3\times 10^{23}$\\
           & 0.97 & 1.00 & 1.310 & $-$3.372 & $-$4.682 & $6.3\times 10^{23}$\\
           & 1.01 & 1.25 & 1.275 & $-$3.124 & $-$3.935 & $3.6\times 10^{23}$\\
\hline
$^{128}$Te & 0.86 ($\beta^-$ decay) & 1.25 & 1.939 & $-$4.276 & 
$-$5.509 & $5.2\times 10^{24}$\\
           & 0.89 & 1.00 & 1.866 & $-$3.975 & $-$5.841 & $1.1\times 10^{25}$\\
           & 0.92 & 1.25 & 1.792 & $-$3.650 & $-$4.790 & $6.9\times 10^{24}$\\
\hline
$^{130}$Te & 0.84 & 1.00 & 1.699 & $-$3.743 & $-$5.442 & $5.3\times 10^{23}$\\
           & 0.90 & 1.25 & 1.575 & $-$3.219 & $-$4.221 & $3.5\times 10^{23}$\\
\hline
$^{136}$Xe & 0.74 & 1.00 & 1.104 & $-$2.615 & $-$3.719 & $1.1\times 10^{24}$\\
\end{tabular}
\end{ruledtabular}
\end{table}

Our results for $0\nu\beta\beta$ nuclear matrix elements disagree with
those of \cite{ROD06} and \cite{SIM99}. In fact, just recently 
\cite{ROD07} the T\"ubingen-Caltec collaboration has corrected their 
results for a coding error in their computer program. The results of
the Erratum \cite{ROD07} agree nicely with our results for the Jastrow
corrected nuclear matrix elements. This means that one can safely say 
that the Jastrow short-range correlations reduce the values of 
matrix elements some 25\%--40\%. On the other hand, for the UCOM we
obtain only a 4\%--16\% reduction. For this reason our present,
UCOM corrected matrix elements are larger than the Jastrow corrected 
ones. Such differences give rise to big differences in the 
predicted $0\nu\beta\beta$ half-lives 
for a given value of the effective neutrino mass $\langle m_{\nu}\rangle$.
This invariably affects the sensitivity estimates for the presently
running and planned double beta experiments.

\section{Summary and conclusions}

We have calculated the $0\nu\beta\beta$ nuclear matrix elements for the
decays of $^{96}$Zr, $^{100}$Mo, $^{116}$Cd,
$^{128}$Te, $^{130}$Te and $^{136}$Xe by using the
proton-neutron quasiparticle random-phase approximation with 
realistic two-body interactions in realistic single-particle spaces.
We have corrected the bare matrix elements for higher-order terms of 
the nucleonic weak currents, for the nucleon's
finite-size and for the nucleon-nucleon short-range
correlations. The short-range correlations have been included
by using the unitary correlation operator formalism. This method is superior
to the rudimentary Jastrow procedure, traditionally adopted for the 
$0\nu\beta\beta$ calculations. 

The UCOM reduces the magnitudes of the matrix elements less than the
Jastrow procedure. This leads to larger matrix elements and shorter 
$0\nu\beta\beta$ half-lives as compared to some recent calculations
quoted in the literature. This has a notable influence on the
estimated neutrino-mass sensitivities of the presently running and 
future double beta experiments.

\begin{acknowledgments}
This work has been partially supported by the 
Academy of Finland under the Finnish Centre of Excellence Programme 
2006-2011 (Nuclear and Accelerator Based Programme at JYFL). We thank
also the EU ILIAS project under the contract RII3-CT-2004-506222.
\end{acknowledgments}

\end{document}